# Title: Electro-optic frequency combs carrying orbital angular momentum

**Authors:** Jinze He[1], Xingyu Jia[1], Bingyan Wei[2]*, Guanhao Wu[1]* and Yang Li[1]*

[1]State Key Laboratory of Precision Measurement Technology and Instruments, Department of Precision Instrument, Tsinghua University, Beijing 100084, China

[2]Key Laboratory of Light Field Manipulation and Information Acquisition, Ministry of Industry and Information Technology, and Shaanxi Key Laboratory of Optical Information Technology, School of Physical Science and Technology, Northwestern Polytechnical University, Xi'an 710129, China

*yli9003@tsinghua.edu.cn

**Abstract:** To date, orbital angular momentum (OAM) and electro-optic optical frequency combs (EOFCs) are two distinct fields of research without too much association. Herein, we generated EOFCs with an OAM on each comb line by applying electro-optic phase modulation to the OAM beam. We verified that the OAM characteristic of the sidebands is consistent with that of the pump light by checking the phase distribution of OAM-EOFC. We also demonstrated the change in the interference contrast and dispersion-induced spiral fringe rotation under different electro-optic modulation strengths. Leveraging the optical rotational Doppler effect of OAM beam and inter-mode beat signal of EOFCs, we achieved rotational speed measurement and absolute distance measurement using a single OAM-EOFC. We also demonstrated, for the first time, the simultaneous measurement of rotational speed and absolute distance of a rough object by tuning the repetition rate of EOFCs. The error of rotational speed measurement is as small as 0.44% and the measured average distance is 3.076 m with a standard deviation of 2.3 mm. Our study bridges two distinct research fields—EOFCs and OAM—opening the door to various fundamental research avenues and applications, including large-capacity optical communications, high-security optical encryption, multi-dimensional photon entanglement, and synthetic dimensions.

- **Introduction**

Angular momentum, a significant feature of light, includes circular polarization-dependent spin angular momentum and orbital angular momentum (OAM) associated with the spatial distribution of the light field[1]. Light-carrying OAM is linked to a helical phase profile $\exp(il\theta)$, where $l$ is the topological charge that can be an arbitrary integer and $\theta$ is the azimuthal angle in the cylindrical coordinate. Owing to their unique spiral phases, inherent orthogonality, and ideal infinite OAM states, OAM beams are particularly appealing in classical and quantum information processing[2,3], nonlinear optics[4,5], metrology[6], and microscopy[7]. Among these areas of applications, the study of OAM in nonlinear media facilitates the exploration of nonlinear dynamics of spatially structured light. So far, nonlinear effects based on OAM beams have been demonstrated in several conventional nonlinear processes, including second-harmonic generation[5,8], four-wave mixing[9], high-harmonic generation[10], and parametric down-conversion[4]. Recently, this research area was extended to cutting-edge physics, such as harmonic spin–OAM cascade[11], second-harmonic generation of spatiotemporal OAM beams[12], and nonlinear beam shaping based on nanomaterials[13].

Most of the existing nonlinear OAM-frequency conversion processes only undergo a single nonlinear process with a few frequency components due to the limitations imposed by phase matching conditions and pump power. Broadband and wavelength-tunable OAM light sources are still in great demand for numerous applications, such as large-capacity optical communications and quantum key distributions. Although there have been several approaches for wavelength-tunable OAM generation by utilizing acoustically-induced fiber grating[14], micro-electro-mechanical filter system[15], and dual-off axis pumped Yb:CaGdAlO$_4$ laser system[16], they still meet the challenges of restricted bandwidth and complicated systems. Thus, it is imperative to generate broadband and wavelength-tunable OAM beams via a simple system. Optical frequency combs (OFCs)[17] can generate a series of discrete, equally spaced frequency comb lines over a wide bandwidth via relatively simple systems, demonstrating the potential to achieve broadband and wavelength-tunable OAM light sources which can enable the simultaneous measurement of distance[18] and rotational speed[6]. Recently, several works have been reported[19,20]. In ref.[19], mode-locked laser pulses were prepared with OAM using a spatial light modulator (SLM). However, the mode purity of each comb line cannot be satisfied due to the intrinsic phase modulation characteristic of SLM. Another route is integrated microring-based vortex soliton microcomb[20], which is limited by low radiant power and low tunability. In contrast to mode-locked laser frequency comb and $\chi^{(3)}$ nonlinearity-based soliton microcomb[17], $\chi^{(2)}$ nonlinearity-based electro-optic (EO) frequency comb (EOFC)[21,22] features inherent stable phase between the comb lines and tunable repetition rate.

Herein, we demonstrated, for the first time, the generation of EOFC with an OAM on each frequency comb line by leveraging EO modulation. The OAM characteristics of EOFC were verified by observing the phase distribution of OAM-EOFC and the interference patterns between the combs carrying OAM and Gaussian combs. Moreover, to further verify the OAM property of each comb, we observed the change in the interference contrast and dispersion-induced spiral fringe rotation under different EO modulation strengths. Leveraging the rotational Doppler effect[6] of OAM beam and inter-mode beat signal of EOFCs, we demonstrated rotational speed measurement and absolute distance measurement using a single OAM-EOFC. The measurement error of rotational speed is less than 0.51% and the measured distance of 5.376 m is with a standard deviation (STD) of 7.7 mm. Moreover, we demonstrated, for the first time, the simultaneous rotational speed measurement and absolute distance measurement of a rough object by tuning the repetition rate of EOFC. The error of rotational speed measurement is as small as 0.44% and the measured average distance of 3.076 m is with a STD of 2.3 mm.

- **Results**

**Design and principle**

The physical process of generating EOFCs with OAM can be divided into two steps, as shown in Fig. 1a and 1b. We first generate the OAM beam using an OAM generator, here a liquid crystal q-plate[23,24] with a spatially variant optical axis distribution classified by the value $q$ ($q = 1/2$) was used. The q-plate can convert the input circularly polarized beam $A_0\exp(i\omega_c t)$ ($A_0$ denotes the amplitude of the light wave and $\omega_c$ is the center frequency) to an OAM beam $A_0\exp(i\omega_c t + il\theta)$ with a topological charge $l = 2q$ in a circular polarization state with opposite handedness to the input beam. In the second step, we transform the generated circularly polarized OAM beam into a linearly polarized state and then inject it into a lithium niobite free-space EO phase modulator (EO-PM-R-20-C3, Thorlabs) with a resonant frequency of 20 MHz driven by a microwave source $A_m\exp(i\omega_m t + \varphi_m)$ ($A_m$ and $\omega_m$ are the amplitude and frequency of the microwave, respectively, while $\varphi_m$ is the phase noise originating from the microwave source). The amplitude of the modulated OAM beam is given by (see Methods)

$$A(\omega) = A_0 \sum_{n=-\infty}^{\infty} J_n(\beta)\delta(\omega - n\omega_m - \omega_c)\exp(il\theta + in\varphi_m), \quad (1)$$

where $\beta$ is the phase modulation index, $J_n$ is the $n$th-order Bessel function of the first kind, and $\delta$ represents the Delta function. The phase modulation corresponds to the generation of EOFCs with a repetition rate of $\omega_m$, with the intensity of comb lines following the form of a Bessel function. Meanwhile, the OAM information carried by the pump light is duplicated to sidebands without any distortion. Figure 1c shows the measured intensity profile and spectrum of OAM-EOFC.

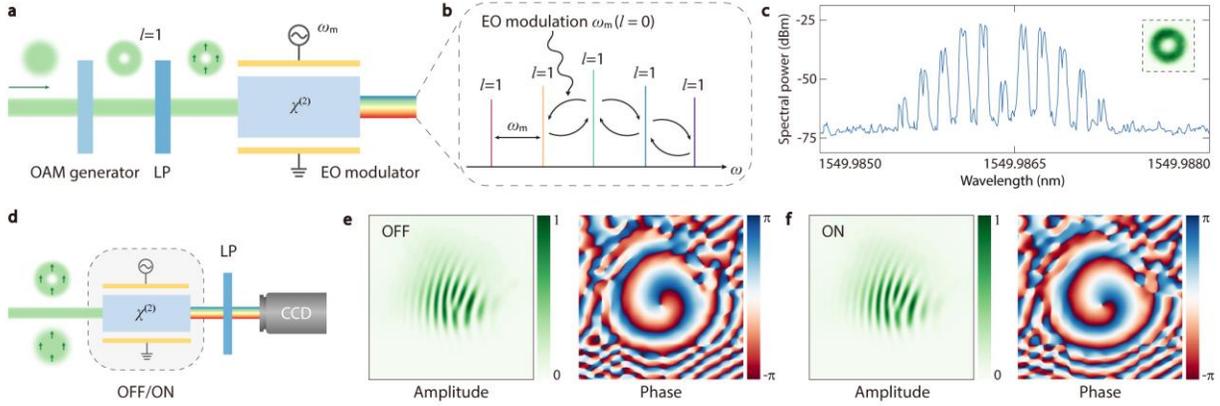

**Fig. 1| Schematics and results of the experiment for generating and characterizing EOFCs carrying OAM. a,** Schematic of OAM-EOFCs generation in two steps: incident wave is converted into the OAM beam via an OAM generator; the OAM beam is modulated by a $\chi^{(2)}$ nonlinearity-based EO modulator to generate OAM-EOFCs. LP, linear polarizer. **b,** Illustration of nonlinear EO modulation as a cascade process of SFG and DFG. OAM characteristics carried by the sidebands are consistent with those of the pump light. The repetition rate of EOFCs coincides with the modulation frequency $\omega_m$. **c,** Measured spectrum of OAM-EOFC. Inset: measured intensity profile of OAM-EOFC. **d,** Schematic for characterizing the OAM properties of sidebands. **e, f,** Measured forklike fringe and retrieved phase distribution when the EO modulator is off and on, respectively.

An alternative perspective considers the nonlinear phase modulation as a cascade process of sum-frequency generation (SFG) and difference-frequency generation (DFG). In this process, the pump light interacts with the microwave with a topological charge $l = 0$ in a $\chi^{(2)}$ nonlinear medium, generating the first-order sidebands (Fig. 1b). Subsequently, pump light and first-order sidebands couple with the microwave in turn generate new sidebands and so forth. The OAM properties of the generated wave are consistent with those of the pump light in the SFG and DFG processes owing to the determinacy of the phase[4], illustrating the consistency of the OAM carried by the sidebands with that carried by the pump light.

To characterize the optical properties of the frequency combs with the OAM and overcome the limitation of the low repetition rate of free-space EO modulators (20 MHz), we designed a set of control experiments (Fig. 1d). When the modulator is off, the interference between the unmodulated monochromatic OAM beam and the monochromatic reference Gaussian beam yields forklike fringes with high contrast and helical phase structure (Fig. 1e). When the modulator is on, both the signal OAM beam and the reference Gaussian beam are modulated, yielding a series of sidebands. The interference between the signal OAM-EOFC and the reference Gaussian EOFC also yields helical phase structure and interference fringes whose contrast is as high as that of the interference between the unmodulated monochromatic OAM beam and the monochromatic reference Gaussian beam (Fig. 1f). Such a phenomenon can only occur when each comb line of the signal OAM-EOFC possesses OAM information.

## Dispersion effect of OAM-EOFCs

To further verify that each comb line does carry OAM, we investigated the variation in the interference spiral fringe between the OAM-EOFC and the Gaussian EOFC by adjusting the power ratio between the center band and sidebands. Similar to the Gaussian beam, the phase of the OAM beam in the propagation direction is related to the propagation distance and wavelength. Therefore, the phase of the interference spiral fringes along the propagation direction changes with the propagation distance and wavelength. In the case of good collimation, the propagation phase of the spiral fringes formed by the $n$th-order comb line varies as $\exp(ik_n\Delta L + in\Delta\varphi)$, where $k_n$ is the wavevector of the $n$th-order comb line, $\Delta L$ is the optical path difference between two arms, and $\Delta\varphi$ is the phase noise difference introduced by the microwave signals in two arms.

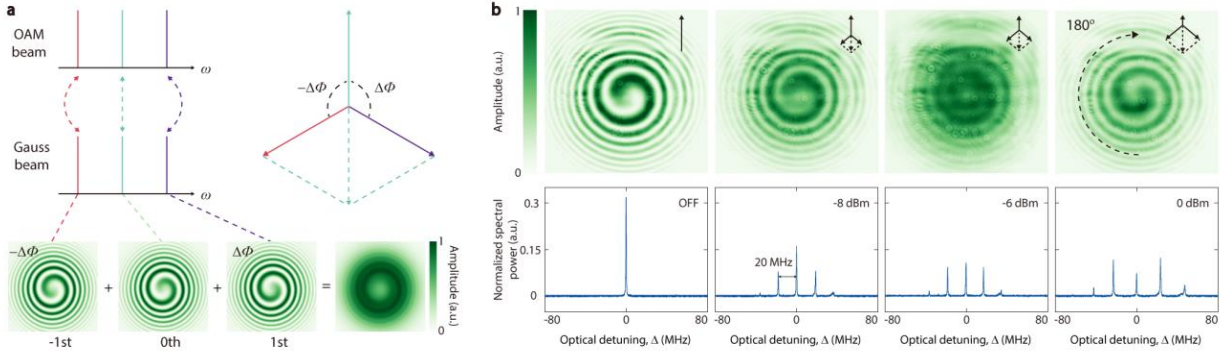

**Fig. 2| Schematic, measured interference fringes, and spectra for verifying the dispersion effect of OAM-EOFCs. a,** Schematic of the dispersion effect of OAM-EOFCs. **b,** Top: Measured interference patterns between OAM-EOFCs and EOFCs without OAM under different modulation strengths. The lengths of the white arrows represent the powers of the center band and the ±1st-order sidebands. Bottom: Spectra of OAM-EOFCs under different modulation strengths measured by the scanning Fabry-Perot interferometer (SA30-144, Thorlabs).

As shown in Fig. 2a, we considered a pair of coherent OAM-EOFC and Gaussian EOFC with the same intensity distribution (the same modulation strength); the phase difference between the fringe at the center band and the fringe at the dominant +1st-order sideband (−1st-order sideband) can be expressed as $\Delta\phi = \Delta k \Delta L + \Delta\varphi$ ($\Delta\phi = -\Delta k \Delta L - \Delta\varphi$), where $\Delta k$ is the absolute value of the wavevector difference between the center band and the +1st-order sideband (−1st-order sideband). The relationship between the center band and ±1st-order sidebands can be interpreted via a vector analysis (top right of Fig. 2a), in which each vector corresponds to the interference between the center or the sideband with OAM and that without OAM, i. e., each vector corresponds to the interference spiral fringe at the bottom of Fig. 2a. Because of dispersion, the spiral fringes formed by the ±1st-order sidebands at a certain $\Delta L$ produce the same amount of phase increment $\Delta\phi$ and decrement $-\Delta\phi$, respectively, corresponding to the clockwise and counterclockwise rotations in the spiral fringes. If $\pi/2 < \Delta\phi < 3\pi/2$, the spiral fringes formed by the superposition of the ±1st-order sidebands are rotated 180° relative to that of the center band, indicating that the spiral fringes of the superposition of the ±1st-order sidebands and the center band are complementary to each other.

We confirmed our analysis of the spiral fringe between Gaussian EOFC and OAM-EOFC by experimentally observing the spiral fringes under various modulation strengths Fig. 2b. As shown in the bottom panel of Fig. 2b, as the modulation strength increases, the ±1st-order sideband power increases. When the power of the vector superposition of the two sidebands is lower than that of the center band, the contrast of the fringe formed by the interference between OAM-EOFC and Gaussian EOFC becomes weaker than that of the fringe formed by the monochromatic OAM beam and the monochromatic Gaussian beam (left two panels of the first row of Fig. 2b). When the power of the vector superposition of the two sidebands is equal to that of the center band, the superimposed spiral fringe of the ±1st-order sidebands is completely complementary to the fringe of the center band, minimizing the contrast of the overall interference fringe (theory: bottom of Fig. 2a; experiment: the third panel of the first row of Fig. 2b). Further increase in the power of the two sidebands can result in the domination of the center band by the positive first-order sidebands, leading to an increase in fringe contrast and reversal of the overall fringe (last panel of the first row of Fig. 2b). Dispersion-induced fringe variations can serve as the second proof of the EOFCs carrying OAM.

**Rotational speed and distance measurement using a single OAM-EOFC**

OAM-EOFC features both the rotational Doppler effect[6], which is intrinsic to the OAM beam (see Methods), and the inter-mode beat signal of EOFC, thus enabling measurement of rotational speed and absolute distance. Here, we used the superposition of two OAM beams with opposite value of $l$ ($l = \pm 20$) as light source (inset of Fig. 3a) to measure the rotational speed $\Omega$ of a rotor with rough surface from which the scattered light was captured by a photodetector (PD1). Figures 3b and 3c show the measured frequency-domain signals and the corresponding rotational speeds when the rotating object was set at different rotational speeds. The difference between the experimental result and the reference is less than 0.51%. We also validated that the modulation strength does not influence the intensity of frequency shift induced by optical rotational Doppler effect (Fig. 3d), reconfirming the consistency of the OAM carried by the sidebands with that carried by the pump light.

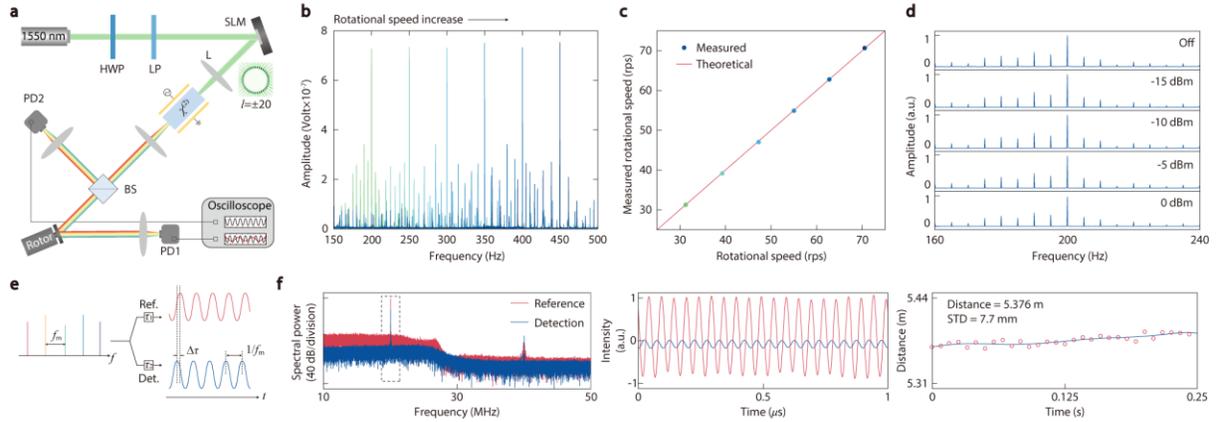

**Fig. 3| Setup, schematic, and experimental results for the measurements of rotational speed and distance using a single OAM-EOFC. a,** Setup for measurement of rotational speed and distance using a single OAM-EOFC. Inset: the intensity profile of the superposition of two OAM beams with $l = \pm 20$ generated using a SLM. HWP, half-wave plate; L, lens; BS, beam splitter; PD, photodetector. **b,** Measured frequency-domain signals under different rotational speeds. **c,** Measured rotational speed versus the reference rotational speed of the object. **d,** Measured frequency-domain signals under different modulation strengths at a rotational speed $\Omega$ of 31.4 rad/s. **e,** Schematic of the distance measurement using a single OAM-EOFC. **f,** Measured spectral (left) and temporal (middle) signals in reference and detection arms. The deduced average absolute distance is 5.376 m with a STD of 7.7 mm in 0.25 s (right).

Figure 3e illustrates the principle of absolute distance measurement using a single OAM-EOFC. In time domain, the inter-mode beat of an OAM-EOFC with a repetition rate of $f_m$ can form a waveform with a period of $1/f_m$. Such a waveform in the reference and detection paths show time delays of $\tau_1$ and $\tau_2$, respectively, enabling deducing the difference between reference and detection paths via the time difference $\Delta\tau = \tau_2 - \tau_1$. In our experiment, we first detected the time-domain signals of reference and detection arms (Fig. 3f, middle) using PD1 and PD2, respectively (Fig. 3a). Based on these time-domain signals, we obtained the frequency-domain signals of reference and detection arms via Fourier transform (Fig. 3f, left), leading to a measured average absolute distance of 5.376 m with a STD of 7.7 mm in 0.25 s (Fig. 3f, right).

Based on this method, we cannot simultaneously measure the rotational speed and the absolute distance of a rough object. This is due to the time-varying noise induced by the uneven reflections of the rough surface, the mechanical jitter of the rotor, and the limited active area of the photodetector. Such a time-varying noise leads to jitter of signal position in time domain and in turn prevents the absolute distance measurement of a rough surface while rotating (see Methods). To avoid the effect of rotating rough surface on distance measurement, a previous work used a SLM to simulate a rotating target for simultaneous measurement of rotational speed and absolute distance. To accurately measure the absolute distance of a rotating rough object, we should consider the frequency-domain distance measurement method.

**Simultaneous measurement of rotational speed and distance of rough surface**

By achieving absolute distance measurement via repetition rate modulated frequency comb (RRMFC)[18], OAM-EOFC can simultaneously measure the rotational speed and distance of a rotating rough surface. RRMFC is enabled by a unique feature of EOFC — the repetition rate can be tuned over a large range in a high speed — in comparison to mode-locked laser and Kerr comb. The principle of RRMFC is shown in Fig. 4a. While the modulation frequency of an EOFC is linearly swept over a range of $\Delta\omega_m$ with sawtooth waveform, the $\pm n$th-order ($n$=1, 2, 3…) sidebands will sweep in a range of $n\Delta\omega_m$. For each pair of these sidebands, the time delay introduced by the optical path difference between the reference and detection arms produces a low-frequency beat, whose frequency is proportional to the order of the sidebands. Hence, by measuring the beat frequency of each pair of sidebands, we can calculate the distance to be measured (see Methods). In contrast to the time-domain distance measurement method, phase noise induced by the rough surface does not affect the frequencies of sidebands. Therefore, OAM-EOFC can accurately measure the distance of a rotating rough object using RRMFC, enabling the simultaneous measurement of rotational speed and distance. In this measurement, to modulate the OAM-EOFC's repetition rate over a broad bandwidth (4 GHz) with a high rate (4 THz/s), we replaced the free-space EO modulator with a fiber-based EO modulator.

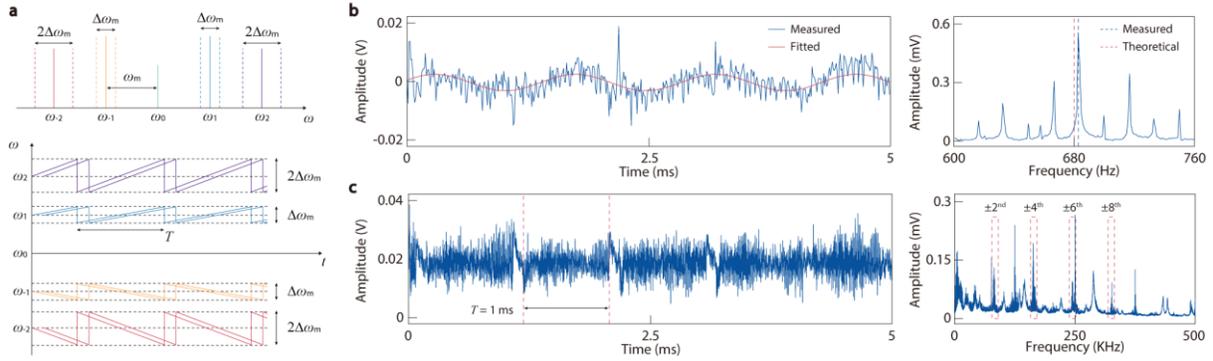

**Fig. 4| Schematic and measured results of simultaneous measurement of rotational speed and distance using OAM-EOFC. a,** Schematic of the distance measurement by tuning the OAM-EOFC's repetition rate $\omega_m$ with sawtooth waveform having a period of $T$ and bandwidth of $\Delta\omega_m$. **b,** Measured time-domain interferogram (left) and the corresponding spectrum (right) for measuring the rotational speed. **c,** Measured time-domain interferogram (left) and the corresponding spectrum (right) for measuring the distance. The sweeping period $T$ of the modulation signal is 1 ms. The spectrum highlights the beat frequencies corresponding to the $\pm 2^{nd}$, $\pm 4^{th}$, $\pm 6^{th}$, $\pm 8^{th}$-order sidebands of EOFC.

Simultaneous time-domain measurement results of rotational speed and distance are shown in the left panels of Fig. 4b and Fig. 4c, respectively. In this measurement, to improve the signal-to-noise ratio and to separate the signals for measuring rotational speed and distance, we used two PDs to detect the signals for rotational speed measurement and distance measurement. The frequency-domain signal in Fig. 4b illustrates a modulation frequency of 683 Hz. Such a frequency is induced by the rotation of the object in a speed $\Omega$ of 106.8 rad/s, showing an error of 0.44%. And, the frequency-domain signal in Fig. 4c shows a series of beat frequencies induced by the different orders of sidebands of EOFC. The amplitudes of these beat frequencies are distributed according to a Bessel function, which is determined by the phase modulation process of EOFC. To better illustrate the sequence of these beat frequencies, we show the frequency-domain signal over time in Fig. 5a. To obtain the absolute distance, the Fourier transform is performed every 10 sweeping periods of the modulation signal. In this way, we obtained about 150 ranging results in 1.5 s, showing an average distance of 3.076 m with a STD of 2.3 mm (Fig. 5b). Due to the small displacements introduced by the jitter of the motor, the measured distance gradually increases as time increases.

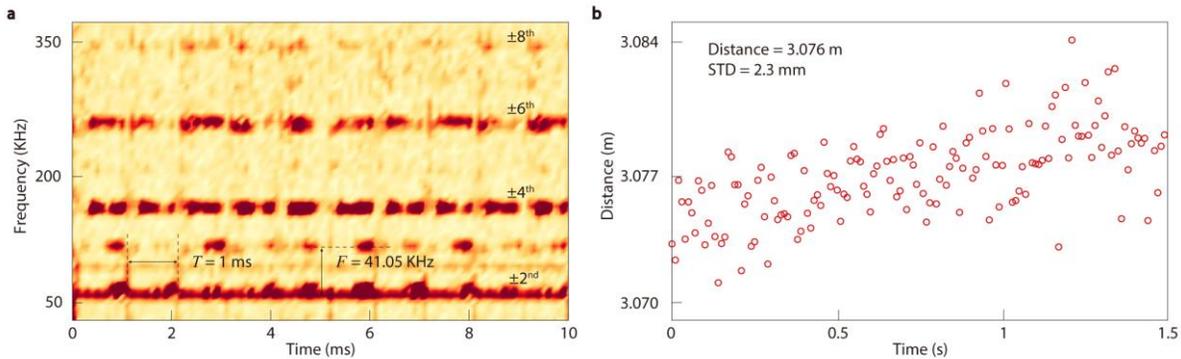

**Fig. 5| Experimental results of the distance measurement by sweeping the repetition rate of OAM-EOFC. a,** Measured spectrum as a function of time. The sweeping period $T$ of the modulation signal is 1ms. The average frequency difference between adjacent beat frequencies $F$ is 41.05 KHz. **b,** Measured absolute distances as a function of time.

- **Discussion**

We reported the generation of EOFC carrying OAM on each comb line. OAM properties of the EOFCs were verified by observing the contrast of the overall interference fringes between the EOFC carrying the OAM and Gaussian beams, as well as the phase distribution of OAM-EOFC. We further verified the OAM characteristic of each comb line by analyzing the dispersion of OAM-EOFC under different modulation strengths. OAM-EOFCs feature the advantages of simplicity, deterministic nature, and high efficiency. Moreover, the tunable center frequency and the repetition rate of EOFC can further increase the flexibility of OAM-EOFCs.

Using a single OAM-EOFC, we achieved rotational speed measurement and absolute distance measurement. The uncertainty in measured rotation speed is less than 0.51% while the STD of measured distance is 7.7 mm for an absolute distance of 5.376 m. By sweeping the repetition rate of EOFC, we demonstrated, for the first time, the simultaneous measurement of rotational speed and absolute distance of a rotating rough object. The error of rotational speed measurement is down to 0.44% while STD of measured distance is only 2.289 mm for an absolute distance of 3.076 m.

The proposed principle of OAM-EOFC could be generalized in various embodiments, including broadband OAM-OFCs based on mode-locked lasers as well as resonator-based EOFC and Kerr combs, in which EOFC and Kerr combs may enable the integration of OAM-OFCs on a photonic chip. Moreover, OAM-EOFCs can serve as a platform to facilitate the improvement of classical and quantum information science, including large-capacity optical communications[25], multi-dimensional photon entanglement[3], and high-security optical encryption based on OAM[26].

- **Methods**

**OAM-EOFC generation.** The 1550 nm continuous-wave (CW) light $A(t)$ passes through an OAM generator can be briefly written as $A_0\exp(i\omega_c t + il\theta)$, where $A_0$ is the amplitude of light, $\omega_c$ is the center frequency. After being modulated by a microwave with amplitude $A_m$ and frequency $\omega_m$, the modulated optical field can be expressed as

$$A(t) = A_0\exp(i\omega_c t + il\theta + \beta\exp(i\omega_m t + \varphi_m)), \qquad (2)$$

where $\beta = \dfrac{\omega_m n^3 r_{33}}{2c} A_m L$ denotes the coefficient of phase modulation, c is the speed of light and $n$, $r_{33}$, $L$ are the refractive index, EO coefficient and modulation length of lithium niobate crystal, respectively. By taking the Fourier transform, the modulated optical field in frequency domain is given by

$$A(\omega) = A_0 \sum_{n=-\infty}^{\infty} J_n(\beta)\delta(\omega - n\omega_m - \omega_c)\exp(il\theta + in\varphi_m), \qquad (3)$$

where $J_n$ is the $n$th-order Bessel function of the first kind. This expression indicates that the phase noise of the microwave source increases linearly with the index line number. In experiment, the free-space EO modulator is driven by a modulation signal from a microwave source (E8257D, Keysight) that is amplified by a microwave power amplifier (ZHL-1-2W+, Mini-Circuits).

**Rotational speed measurement of a rough surface using OAM-EOFC.** A rough surface with an uniform reflectivity can be considered as a pure phase modulator whose phase is related to the polar coordinates $r$ and $\theta$, which can be expanded in Fourier expansion form as[27]

$$M(r,\theta) = \sum_p B_p(r)\exp(ip\theta), \qquad (4)$$

where $B_p(r)$ is the modulation amplitude of each order and can be normalized as $\sum_p |\int B_p(r)dr|^2 = 1$. When the OAM-EOFC is incident onto a rough surface with a rotational speed of $\Omega$, according to the rotational Doppler effect, the scattered OAM-EOFC can be expressed as

$$A_{out}(r,\theta,\Omega,t) = A_0\exp(i\omega_c t + \beta\exp(i\omega_m t + \varphi_m))\sum_p B_p(r)\exp(i(p-l)\theta + p\Omega t). \qquad (5)$$

Here, the negative sign is introduced by reflection. If the illumination beam is comprised by the superposition of two OAM beams with opposite values of $\pm l$, the scattered light from the rough surface will produce a beat with a modulation frequency of $\Omega l/\pi$ which can be captured by a photodetector.

**Simultaneous measurement of rotational speed and distance using OAM-EOFC.** In the previous section, we assumed that the modulation amplitude $B_p(r)$ of the rough surface is independent of time. However, in experiments, the modulation amplitude of the rough surface is a time-dependent variable ($B_p(r,t)$) due to the uneven reflections of the rough surface, the mechanical jitter of the rotor, also the limited active area of the photodetector. Then, the temporal variation of the modulation amplitude introduces phase noise to the time-domain signals of the EOFC. Consequently, we cannot obtain the distance information from the relative position between the measured time-domain inter-mode beat signals from the reference and detection arms. However, in the frequency domain, the time-domain phase noise introduced by the rough surface does not change the frequency of the sidebands of OAM-EOFC, as shown below.

$$A_{out}(r,\theta,\Omega,\omega) = A_0 \sum_{n=-\infty}^{\infty} \sum_p J_n(\beta) B_p(r,t)\delta(\omega - n\omega_m - \omega_c)\exp(in\varphi_m)\exp(i(p-l)\theta + p\Omega t). \qquad (6)$$

Here we leverage the tunable repetition rate of EOFC to achieve RRMFC ranging by sweeping the frequency of modulation signal. The RRMFC scattered by the rough surface can be expressed as

$$A_{out}(r,\theta,\Omega,t) = A_0\exp(i\omega_c t + \beta\exp(i(\omega_m + \dot{\omega}_m t)t + \varphi_m))\sum_p B_p(r,t)\exp(i(p-l)\theta + p\Omega t), \qquad (7)$$

where $\dot{\omega}_m = \Delta\omega_m/T$ is the sweeping speed of repetition rate. We can get the beat frequency generated by each order of comb line through the interferogram between reference arm and detection arm with a distance to be measured of $D$

$$f(n) = n\frac{\Delta\omega_m}{2\pi T}\frac{D}{c}. \tag{8}$$

Therefore, we can derive the distance from the beat frequency of the *n*th-order comb line via $D = \frac{2\pi c T f(n)}{n\Delta\omega_m}$.

## Data availability

The data that support the findings of this study are included in the published article. Source data are provided with this paper.

**Acknowledgments**

The authors would like to acknowledge Z. Y. Sun for inspiring the authors in conceiving the basic idea of this work. This work was supported by the National Key Research and Development Program of China (2021YFA1401000, 2021YFB2801600), National Natural Science Foundation of China (62075114, 12074313, 12374279), and Beijing Natural Science Foundation (4212050). This work was supported by the Center of High-Performance Computing, Tsinghua University.

**Author contributions**

Y. Li and J. Z. He conceived the basic idea for this work. J. Z. He and B. Y. Wei carried out the simulations. B. Y. Wei performed the fabrication. J. Z. He and X. Y. Jia carried out the measurements and analyzed the experimental results. B. Y. Wei, G. H. Wu and Y. Li supervised the research and the development of the manuscript. J. Z. He wrote the first draft of the manuscript; all authors subsequently took part in the revision process and approved the final copy of the manuscript.

**Competing interests**

The authors declare no competing interests.